\documentclass[pra,twocolumn,a4paper,showpacs,superscriptaddress,nofootinbib]{revtex4}
\usepackage{amsmath}
\usepackage{amsfonts}
\usepackage{graphicx}

\begin{document}

\title{Dynamics of Bose-Einstein Condensates in 
  One-Dimensional Optical Lattices in the Presence of
  Transverse Resonances}
\author{K.-P. Marzlin}
\affiliation{Department of Physics and Astronomy,
University of Calgary,
2500 University Drive NW, Calgary, Alberta T2N 1N4, Canada}
\affiliation{Fachbereich Physik der Universit\"at Konstanz, 
Fach M674, 78457 Konstanz, Germany}
\author{V.I. Yukalov}
\affiliation{Bogolubov Laboratory of Theoretical Physics, 
Joint Institute for Nuclear Research, Dubna 141980, Russia}
\affiliation{Institut f\"ur Theoretische Physik,
Freie Universit\"at Berlin, Arnimallee 14, D-14195 Berlin, Germany}

\begin{abstract}
The dynamics of Bose-Einstein condensates in the lowest energy band
of a one-dimensional optical lattice is generally disturbed by
the presence of transversally excited resonant states.
We propose an effective one-dimensional theory
which takes these resonant modes into account and derive
variational equations for large-scale dynamics.
Several applications of the theory are discussed and a novel type of 
``triple soliton'' is proposed,
which consists of a superposition of a wavepacket at the upper band edge
and two transversally excited wavepackets which are displaced in
quasi-momentum space.
\end{abstract}


\maketitle

\section{Introduction}

The phenomenon of Bose-Einstein condensation is
a collective effect which relies on the bosonic nature
of the particles alone 
(for reviews see, e.g., Refs.~\cite{orgcit1}-\cite{castin01}). 
Although an interaction between
particles is not needed for the 
corresponding phase transition, its presence has a
substantial influence on the properties of a Bose-Einstein condensate (BEC). 
In this context, solitons are of fundamental interest since they
represent states whose very existence relies on the interaction.

For atomic BECs, bright solitons as well as dark solitons have 
been experimentally demonstrated for atoms with attractive
\cite{khayk02,strecker02} and repulsive interaction 
\cite{burger99,denschlag00}, respectively. The present work is
motivated by the recent observation of gap solitons
in a $^{87}$Rb BEC \cite{markusGapSol03}. Gap solitons are
bright solitons for a BEC with repulsive interaction in an optical
lattice and rely on the negative effective mass around the
upper band edge of the periodic potential.
To create a gap soliton it is necessary
to control the motion of the initial wavepacket in quasi-momentum space
\cite{markusPRL03}. This kind of physical situation has recently
been intensively studied, both theoretically
\cite{kivshar04,kivshar04b,plaja04,konotop04}
and experimentally 
\cite{arimondo03,inguscio04,inguscio04b}.

In the present work we consider the influence of the
transverse confining potential on the dynamics of a BEC
in an one-dimensional optical lattice. We are particularly 
interested in the behaviour around the upper band edge of the lowest
energy band.  In this energy range the transverse confinement
leads to the presence of transversally excited resonant states
which significantly change the stability 
of the BEC \cite{kmh02,modugno04b} and alter its dynamics \cite{modugno04}.
The resonances are important if the transverse
excitation energy is small compared to the modulation depth of the
optical lattice.

Much of the recent research on BEC is concerned with an effectively
one-dimensional situation. Generally this can be achieved if the
transverse excitation energy is large compared to the interaction
energy. This allows a simplified one-dimensional description of 
the dynamics by either projecting the collective wavefunction on
the transverse ground state \cite{orgcit2,castin01} or, more accurately,
by making a Gaussian variational ansatz for the transverse shape 
of the wavefunction \cite{salasnich02,salasnich02b}.
While such an approach gives excellent agreement with a full
three-dimensional theory in absence of transverse resonances
(i.e., around the lower band edge in the case of a 1D optical lattice
\cite{modugno04b}),
it is not suitable to describe a BEC around the upper band edge
\cite{kmh02,modugno04b}.
In this paper we present a generalized
one-dimensional theory by
projecting the collective wavefunction on a superposition of 
longitudinal wavepackets centered around the resonant states.
In Sec.~\ref{sec:problem} we will review the preparation of a
BEC at the upper band edge in order to motivate our particular approach.
In Sec.~\ref{sec:dynEqs} simplified dynamical equations are derived
and compared to previous approaches. In Sec.~\ref{sec:variat}
we further reduce these equations by making a variational ansatz
for the wavefunction.  In Sec.~\ref{sec:sol} we will discuss
several solutions of this system.

\section{Description of the problem}\label{sec:problem}
Very recently,  gap solitons have been 
experimentally observed in a BEC of Rubidium atoms 
\cite{markusGapSol03}. Gap solitons correspond to a wavepacket
of repulsively interacting atoms prepared at the upper band
edge of the lowest band in an optical lattice. 
The process of creating a gap soliton is quite sophisticated
since one has to move the BEC from the ground state,
where it first is created, to the upper
band edge of the optical lattice.
For the purpose of this paper it can be summarized in the 
following way: first,
a BEC is created in the ground state of a 3D harmonic trap
$V_{\mbox{{\scriptsize trap}}} ({\bf x}) = M^2 \omega_\|^2 z^2/2
+ M^2 \omega_\perp^2 (x^2+y^2)/2$, where $M$ is the atomic mass
and $\omega_i$ are the axial and transverse trapping frequencies.
Then a one-dimensional optical lattice of the form
\begin{equation} 
  V_{\mbox{{\scriptsize opt}}} (z,t) = V_0 \cos (2 k_L z + \phi(t))
\label{eq:vopt} \end{equation} 
along the $z$ axis is 
switched on adiabatically and the axial harmonic trap is switched
off ($\omega_\| =0$). Here, $k_L$ is the wavevector of the laser beam
forming the optical lattice. 
At this time the lattice phase $\phi(t)$ is
zero. The BEC is thus prepared as a wavepacket around the lower band
edge of the lowest band of the optical lattice.
Finally, a Bloch oscillation is employed ($\phi(t)$ varying with time)
so that the wavepacket is slowly moving upwards in the energy band 
(so that excitations to higher bands can be neglected) and
eventually reaches the upper band edge. This is an application
of dispersion management for atomic matter waves which is
described in more detail in Ref.~\cite{markusPRL03} and is now
of high experimental interest \cite{arimondo03,inguscio04,inguscio04b}.

To describe the dynamics of a BEC that is manipulated within the
lowest energy band of the lattice, it would be desirable 
to have an effective dynamical equation at hand
which is one-dimensional and based on the effective-mass approximation,
rather than including the full periodic and transverse trapping potentials. 
To derive such an equation
we start from the Gross-Pitaevskii equation (GPE) for a
BEC in a 1D optical lattice and a transverse trapping potential,
\begin{eqnarray} 
  i\hbar \partial_t \psi ({\bf x}, t) &=& \left ( H_\| + H_\perp
    -M a z  \right ) \psi  ({\bf x}, t) 
\nonumber \\ & &
  + \kappa |\psi  ({\bf x}, t)|^2
    \psi  ({\bf x}, t) 
\label{eq:gpe} \end{eqnarray} 
with ${\bf x}_\perp := (x,y)$ and 
\begin{eqnarray} 
  H_\| &=& \frac{p_z^2}{2 M} + V_{\mbox{{\scriptsize opt}}} (z,t) 
  \\ 
  H_\perp &=&  \frac{{\bf p_\perp}^2}{2 M} + 
      V_{\mbox{{\scriptsize trap}}} ({\bf x}_\perp) \; .
\end{eqnarray} 
Here $\psi$ is the collective atomic wavefunction which we assume
to be normalized to one. The interaction parameter is given by
$\kappa := 4 \pi \hbar^2 a_{\mbox{{\scriptsize scatt}}} N/M$ with  
$  a_{\mbox{{\scriptsize scatt}}}$ being the atomic scattering length
and $N$ the number of atoms in the BEC.
We have also included a homogeneous force term which corresponds
to an acceleration $a$ of the atoms. This term is closely related
to the time variation of the lattice phase $\phi (t)$ and
responsible for the generation of Bloch oscillations of the wavepacket.
To avoid exciting atoms to higher bands of the optical lattice,
the acceleration must be small enough so that adiabatic motion
in the lowest band is possible. Throughout the paper we will
assume that this is the case. 
We have omitted a longitudinal
confining potential since our aim is to study
the effects of the transverse dynamics rather than the perturbation
of the longitudinal lattice symmetry. A weak longitudinal potential
could be included by introducing a slow variation of the lattice
parameters \cite{brazhni04}, however.

Being nonlinear and inhomogeneous, Eq.~(\ref{eq:gpe})
is impossible to solve analytically. Even
the numerical simulation of it is time-consuming because of the
necessity to resolve features on the scale of half the laser's wavelength
(which is equal to the period of the lattice).
In addition, it would be desirable to have a description which uses
the (numerically verified) fact that the wavepacket stays localized in
the energy band for a long time if
the modulation depth $V_0$ is sufficiently small. We remark that,
if $V_0$ gets too large, a phase transition
to a spatially localized state which is smeared out over the
lowest energy band takes place instead \cite{trombettoni01}.

To derive such an analytical theory, we employ the observation
that a wavepacket, which is narrowly localized around a certain
quasi-momentum $q_0$ in the lowest energy band, is very broad
and varies slowly in position
space. Let us assume for the moment that no transverse excitations
are produced. Then one can make the ansatz 
\begin{equation} 
  \psi ({\bf x}, t) = B (z,t) \varphi_{q_0} (z) \chi_{0} 
   ({\bf x}_\perp)
\label{eq:simpleAnsatz} \end{equation}  
where $\varphi_{q_0}$ is a quasiperiodic (Bloch) eigenfunction
of  $H_\|$ with quasimomentum $q_0$. The function $\chi_{0}$
denotes the transverse ground state of the trapping potential.
The (dimensionless)
function $B(z,t)$ is an envelope which describes the large scale
features of the wavefunction, whereas the small-scale features
are included in $\varphi_{q_0}$. The basic idea of our approach is 
to average over the small spatial scales and to derive an effective
equation for the large-scale behaviour of the wavefunction, i.e.,
for the envelope $B(z,t)$. 

\section{Effective dynamical equations for resonant modes}
\label{sec:dynEqs}
\begin{figure}
\centerline{
\includegraphics[width=7.5cm]{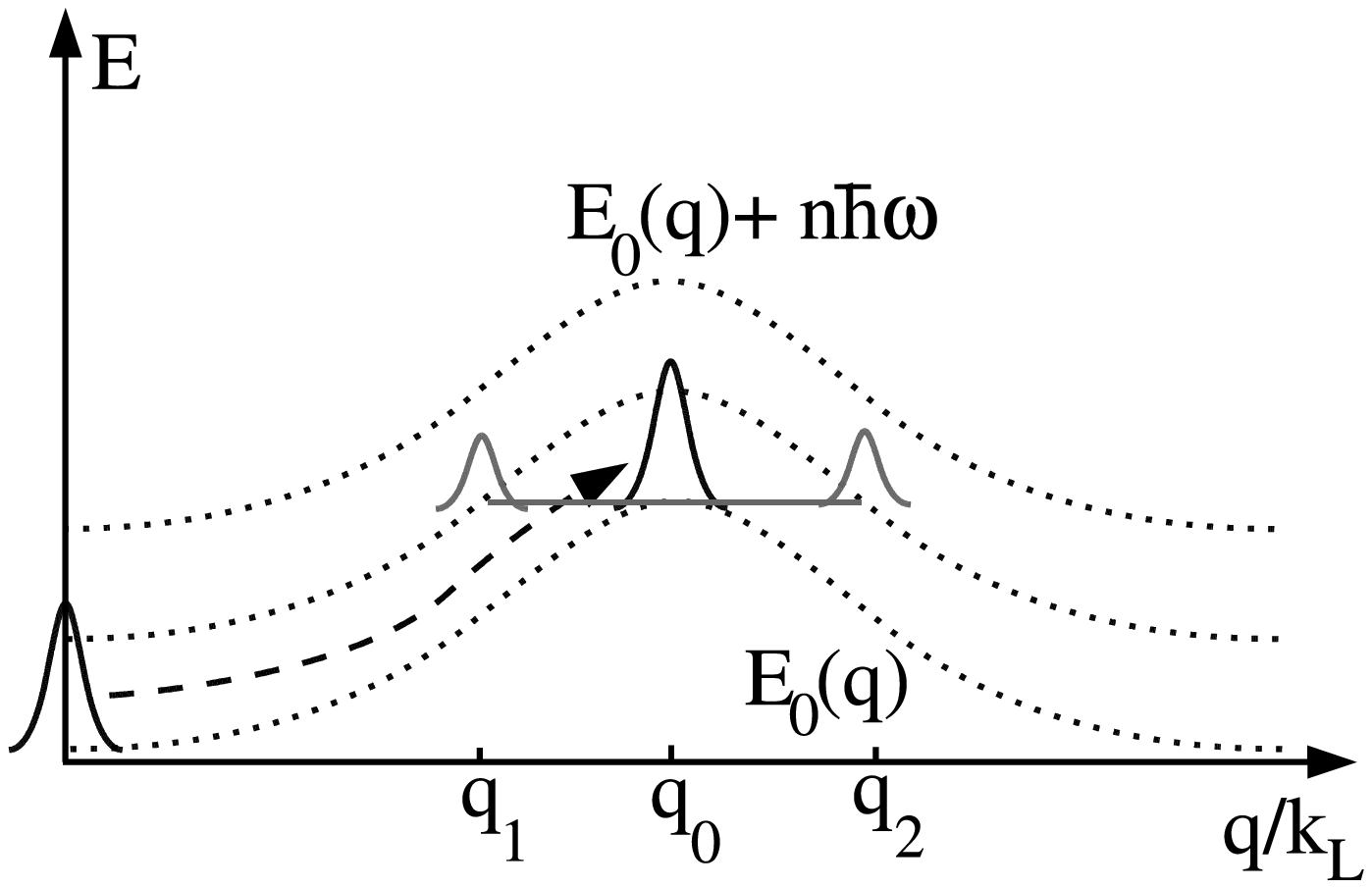}
}
\caption{\label{fig:modeScheme} Scheme of the collective wavepacket's motion
through the lowest energy band. The dotted lines represent the
spectrum of noninteracting atoms in a 1D optical lattice and a
transverse harmonic trap. The lowest of these lines corresponds to
the lowest energy band of the lattice for atoms in the transverse
ground state. The two upper copies of it are transversally excited
atoms in the same band. The BEC is initially prepared as a 
wavepacket around the lower band edge (lower left corner)
and is adiabatically moved
to the upper band edge with quasimomentum $q_0$
(dashed arrow). Around the upper band
edge transversally excited resonances occur at quasimomenta
$q_1$ and $q_2$.}
\end{figure}

Before we can start to derive an effective equation, the ansatz
(\ref{eq:simpleAnsatz}) has to be generalized in two aspects.
First, since our aim includes to describe the adiabatic Bloch
oscillation from the lower to the upper band edge, we cannot
assume that the quasimomentum is fixed, but have to admit
a time-dependent $q_0(t)$. Secondly, we have to take into account 
transverse resonances which appear if $q_0(t)$ is in the
vicinity of the upper band edge (see Ref.~\cite{kmh02} and 
Fig.~\ref{fig:modeScheme}). Numerical investigations
suggest that it is sufficient to include the two nearest resonances
only, because all other resonances are negligibly populated.
Therefore,
the ansatz (\ref{eq:simpleAnsatz}) needs to be modified to
\begin{equation} 
  \psi ({\bf x}, t) = \sum_i B_i (z,t) \varphi_{q_i(t)} (z) 
                      \chi_{n_i} ({\bf x}_\perp),
\label{eq:fullAnsatz} \end{equation} 
where, for our purposes, $i$ runs from 0 to 2, $q_i$ denotes
the quasimomentum around which each of the three modes is
centered, and $n_i$ represents the transverse excitation number
($n_0=0, n_1=n_2=2$ since, by symmetry, only even levels
can be excited \cite{kmh02}). $B_i$ is a slowly varying envelope
function for each of the three modes.

To derive an effective equation for the envelopes, we average
over the small spatial scale set by the lattice length $L_a=\pi/k_L$.
Following standard methods, we introduce an averaging function
$f_{\mbox{{\scriptsize av}}}(z)$ which is slowly varying on the scale
of $L_a$, has a narrow support whose width corresponds to the resolution
of the effective equation, and which is normalized to one, 
$\int f _{\mbox{{\scriptsize av }}}(z) dz=1$. The width of
$f_{\mbox{{\scriptsize av}}}$ should be much smaller than the scale on
which the envelopes $B_i$ are varying. A function $g(z)$ is then
averaged by calculating $\langle\langle g \rangle\rangle(z) := \int dz^\prime 
f_{\mbox{{\scriptsize av}}}(z^\prime) g(z-z^\prime)$.
With this method, the envelopes can be 
extracted from the wavefunction by evaluating
\begin{eqnarray} 
  & &
  \int  dz^\prime f_{\mbox{{\scriptsize av}}}(z^\prime) 
  \int d^2 {\bf x}_\perp\,
   \chi_{n_i}^*({\bf x}_\perp) \varphi_{q_i}^*(z-z^\prime)
   \psi({\bf x}_\perp,z-z^\prime)  
  \nonumber \\ &=&
  \sum_j \delta_{n_i,n_j} 
  \int dz^\prime  f_{\mbox{{\scriptsize av}}}(z^\prime)
  B_j(z-z^\prime ) \varphi_{q_i}^*(z-z^\prime )
   \varphi_{q_j}(z-z^\prime )
  \nonumber \\ &\approx&
  \sum_j B_j(z) \delta_{n_i,n_j} 
  \int dz^\prime  f_{\mbox{{\scriptsize av}}}(z^\prime)
   \varphi_{q_i}^*(z-z^\prime )
   \varphi_{q_j}(z-z^\prime ),
\label{eq:integralEval}\end{eqnarray} 
where we have used that $B_j$ is approximately constant
over the support of $f_{\mbox{{\scriptsize av}}}$ and where the time 
dependence, for brevity, is dropped out. The integral in the last line 
can be evaluated as follows:
for $j=i$ the function $|\varphi_{q_i}|^2$ is periodic with period
$L_a$. Therefore, $\int_0^{L_a} |\varphi_{q_i}|^2 dz = L_a/L$
since the Bloch functions are normalized ($L$ is the quantization
length). Since $f_{\mbox{{\scriptsize av}}}$ is roughly constant
on the scale of $L_a$, we find, by cutting the integral into bits
of length $L_a$,
\begin{eqnarray} 
\int dz^{\prime \prime}   f_{\mbox{{\scriptsize av}}}(z-z^{\prime \prime} )
   |\varphi_{q_i}(z^{\prime \prime} )|^2
  &\approx & \sum_m  f_{\mbox{{\scriptsize av}}}(z- m L_a) \frac{L_a}{L}
  \nonumber \\
  &\approx & \frac{1}{L},
\end{eqnarray} 
since the sum is just the discretized expression for a Riemannian integral
over $f_{\mbox{{\scriptsize av}}}$ with $dz = L_a$. For $j\neq i$,
consider first the case that the width $L_f$ of $f_{\mbox{{\scriptsize av}}}$
is very large, $L_f=L$. Then the integral is simply the scalar product
between the two modes and therefore zero unless $q_j=q_i$.
For sufficiently large $L_f$, the integral is still 
approximately zero if $q_i$ and $q_j$ are not too close to each other,
since the product of the Bloch wavefunctions then oscillates rapidly
and averages to zero.
Assuming that this is the case we find from Eq.~(\ref{eq:integralEval})
\begin{equation}
  \langle \langle 
  \int d^2 {\bf x}_\perp\,
   \chi_{n_i}^* \varphi_{q_i}^*
   \psi \rangle \rangle (z)  = \frac{B_i(z)}{L}.
\end{equation} 
When we apply the same procedure (projecting onto the transverse 
modes and averaging over the longitudinal part) to the GPE 
and insert the ansatz (\ref{eq:fullAnsatz}), we are led to
\begin{eqnarray}   
  i\hbar \dot{B}_i &=& \hbar \omega_\perp (n_i+ \frac{1}{2}) B_i
  + L \sum_j \delta_{n_i,n_j} 
  \langle \langle  \varphi_{q_i}^* H_\|  \varphi_{q_j} B_j 
  \rangle \rangle
  \nonumber \\ & &
  +\kappa \sum_{j,k,l} B_j^* B_k B_l  I^\|_{ij;kl} I^\perp_{ij;kl}
  \nonumber \\ & &
  + (\dot{q}_i - M a) z B_i,
\label{eq:deriv1} \end{eqnarray} 
with the usual interaction mode integrals
\begin{eqnarray} 
  I^\|_{ij;kl} &:=& \int dz \varphi_{q_i}^*  \varphi_{q_j}^*
    \varphi_{q_k}  \varphi_{q_l}
  \\
 I^\perp_{ij;kl} &:=& \int d^2 {\bf x}_\perp
   \chi_{n_i}^*  \chi_{n_j}^* \chi_{n_k}  \chi_{n_l} \; .
\end{eqnarray} 
A dot denotes the derivative with respect to time. In the derivation 
of Eq.~(\ref{eq:deriv1}) we have exploited the fact that
the averaging over the interaction integrals  can be done in much
the same way as for Eq.~(\ref{eq:integralEval}): the averaged
interaction integrals are again either periodic or rapidly
oscillating and therefore do essentially acquire a factor $1/L$,
which we have multiplied out in  Eq.~(\ref{eq:deriv1}).

The last line in Eq.~(\ref{eq:deriv1}) deserves a comment.
The homogeneous potential term $-M a z$ simply survives the
averaging procedure and is a direct consequence of the
corresponding term in Eq.~(\ref{eq:gpe}). The term
proportional to $\dot{q}_i$ arises from the time derivative
on the left-hand side of Eq.~(\ref{eq:gpe}) which includes a term
of the form $\dot{q}_i (\partial_{q_i}\varphi_{q_i}) B_i$.
It is not hard to see that, provided the assumption that the
wavepacket remains in the lowest energy band holds true,
the derivative with respect to the quasi momentum can be
approximated by $ \partial_{q_i}\varphi_{q_i} \approx i z \varphi_{q_i}$.
The term is then of the same form as the homogeneous force
and can be averaged in the same way. It is interesting to
note that in the case of a simple Bloch oscillation caused
by the homogeneous force we have $\dot{q}_i = M a$
so that the linear potential is cancelled. This is nothing
but a different description of the fact that a Bloch oscillation
simply corresponds to a shift of a wavepacket in quasimomentum space,
again under the condition that no higher bands are populated.
This is the case for the main wavepacket in Fig.~\ref{fig:modeScheme} for
which the time dependence of its quasimomentum $q_0$
is simply a consequence of the induced Bloch oscillation.
However, for the modes $q_1$ and $q_2$ the time dependence
of the quasimomenta is determined by a resonance condition
and is not directly related to the Bloch oscillation. Hence,
these two modes are subject to a renormalized homogeneous force.

To perform the averaging over the longitudinal Hamiltonian
in Eq.~(\ref{eq:deriv1}),
we employ the well-known effective-mass method from solid
state theory (see, e.g., Ref.~\cite{hakenQFT}). Using that
$B_j  \varphi_{q_j}$ is narrowly localized around quasi momentum
$q_i$ we can expand this expression in terms of Bloch wave functions
$\varphi_{q_j+\Delta q}$ , which are eigenfunctions of $H_\|$
with eigenvalues $E(q_j+\Delta q)$. The eigenvalue can be expanded
to second order in $\Delta q$, resulting in
\begin{eqnarray} 
  H_\|  \varphi_{q_j} B_j &\approx&
  \int d\Delta q\; \langle \varphi_{q_j +\Delta q}|\varphi_{q_j} B_j \rangle 
 \times \\ & &
  \left ( E(q_j) + v_j\Delta q + \frac{ \Delta q^2}{
   2 M^{\mbox{{\scriptsize  eff}}}_j } \right ) \varphi_{q_j +\Delta q} \; .
\nonumber \end{eqnarray} 
In this equation, we introduced two important physical parameters:
the group velocity $v_j := \partial E(q) /\partial q |_{q=q_j}$
and the effective mass $M^{\mbox{{\scriptsize  eff}}}_j :=
(\partial^2 E(q) /\partial q^2 |_{q=q_j})^{-1}$. Introducing the function
\begin{equation} 
  \tilde{B}_j(z) := \int d\Delta q e^{i \Delta q z} 
  \langle \varphi_{q_j +\Delta q}|\varphi_{q_j} B_j \rangle , 
\label{eq:atilde} \end{equation} 
it is easy to see that the action of $H_\|$ can be expressed as
\begin{eqnarray} 
 (H_\|  \varphi_{q_j} B_j )(z)&\approx& \int \frac{dz^\prime }{2\pi}
 (H_{\mbox{{\scriptsize  eff}},\|}^{(j)} \tilde{B}_j)(z^\prime ) \times
 \nonumber \\ & & 
 \int d\Delta q e^{-i\Delta q z^\prime }  \varphi_{q_j +\Delta q}(z) ,
\end{eqnarray} 
with the effective Hamiltonian
\begin{equation} 
  H_{\mbox{{\scriptsize  eff}},\|}^{(j)} = 
   E(q_j) -i\hbar v_j \partial_z - \frac{\hbar^2 \partial_z^2}{
   2 M^{\mbox{{\scriptsize  eff}}}_j} \; .
\end{equation} 
This allows us to write the averaged Hamiltonian action 
appearing in Eq.~(\ref{eq:deriv1}) in the form
\begin{eqnarray} 
   \langle \langle  \varphi_{q_i}^* H_\|  \varphi_{q_j} B_j 
   \rangle \rangle &=&  
   \int \frac{dz^\prime }{2\pi}
   (H_{\mbox{{\scriptsize  eff}},\|}^{(j)} \tilde{B}_j)(z^\prime )
   \int dz^{\prime \prime} f_{\mbox{{\scriptsize av}}}(z^{\prime \prime}) 
 \times \nonumber \\ & &
  \int d\Delta q \;  e^{i \Delta q (z-z^\prime -z^{\prime \prime })}
   u_{q_i}^*(z-z^{\prime \prime })
 \times \nonumber \\ & & \hspace{1.5cm}
   u_{q_j+\Delta q}(z-z^{\prime \prime }) ,
\end{eqnarray} 
where $u_q$ are the periodic Bloch wavefunctions, 
$\varphi_q (z) = \exp (i q z) u_q(z)$. Because of the averaging,
we are interested in distances $z-z^\prime $ much larger than $L_a$.
In this case, the phase factor in the integral over $\Delta q$ varies
much faster with $\Delta q$ than the periodic Bloch function
$u_{q_j+\Delta q}$. We therefore can replace the latter by $u_{q_j}$.
The integral over $\Delta q$ then becomes, on scales much larger than
$L_a$, the delta function $2\pi \delta (z-z^\prime -z^{\prime \prime })$
and we arrive at
\begin{eqnarray} 
\langle \langle  \varphi_{q_i}^* H_\|  \varphi_{q_j} B_j 
   \rangle \rangle &=&  
   \int dz^{\prime \prime} 
   (H_{\mbox{{\scriptsize  eff}},\|}^{(j)} \tilde{B}_j)(z-z^{\prime \prime } )
    f(z^{\prime \prime }) \times
 \nonumber \\ & &
 \varphi_{q_i}^*(z^{\prime \prime })  \varphi_{q_j}(z^{\prime \prime })
 \nonumber \\ &\approx&
   (H_{\mbox{{\scriptsize  eff}},\|}^{(j)} \tilde{B}_j)(z)   
   \int dz^{\prime \prime} 
    f(z^{\prime \prime })  \times
 \nonumber \\ & &
 \varphi_{q_i}^*(z^{\prime \prime })  \varphi_{q_j}(z^{\prime \prime })
 \nonumber \\ &\approx& \delta_{q_i , q_j}
    \frac{1}{L}  H_{\mbox{{\scriptsize  eff}},\|}^{(j)} \tilde{B}_j   \;.
\end{eqnarray} 

The last step in our derivation of effective equations
for the envelope functions $B_j$ is to show that $B_i$ and 
$\tilde{B}_i$ are, on average, equal. To do so, we first note
that $B_i = L \langle \langle \varphi_{q_i}^*  \varphi_{q_i} B_i 
\rangle \rangle $
since $B_i$ is slowly varying. Inverting Eq.(\ref{eq:atilde}), we
can rewrite this as
\begin{eqnarray} 
    B_i(z) &=& L\int dz^{\prime\prime} f_{\mbox{{\scriptsize av}}} 
    (z^{\prime\prime}) \int d\Delta q \varphi_{q_i}^* (z-z^{\prime\prime})
  \times \nonumber \\ & &
    \varphi_{q_i+\Delta q}(z-z^{\prime\prime}) \int \frac{dz^\prime}{2\pi}
    e^{-i \Delta q z^\prime } \tilde{B}_i(z^\prime)\; .
\end{eqnarray} 
It is then possible to repeat the argument given above for the action
of $ H_\| $. Writing the quasiperiodic Bloch functions $\varphi_q$
in terms of the periodic Bloch functions $u_q$, we again find a 
rapidly oscillating exponential in $\Delta q$ which results in a
spatial delta function for large scales. Integrating this we find
\begin{eqnarray} 
   L \langle \langle |\varphi_{q_i}|^2 B_i \rangle \rangle
   &=& L \langle \langle \tilde{B}_i |u_{q_i}|^2 \rangle \rangle 
   \nonumber \\ &\approx&
   L \tilde{B}_i \langle \langle |u_{q_i}|^2 \rangle \rangle
   \nonumber \\ &=& \tilde{B}_i .
\end{eqnarray}  
Using this identity we find for the effective equation 
describing the large scale dynamics of the envelopes
\begin{eqnarray}   
  i\hbar \dot{B}_i &=& \hbar \omega_\perp (n_i+ \frac{1}{2}) B_i
  +  H_{\mbox{{\scriptsize  eff}},\|}^{(i)}B_i 
  \nonumber \\ & &
  +\kappa \sum_{j,k,l} B_j^* B_k B_l  I^\|_{ij;kl} I^\perp_{ij;kl}
  \nonumber \\ & &
  + (\dot{q}_i - M a) z B_i \; .
\label{eq:effEq} \end{eqnarray} 
For the case of a single wave packet centered around a fixed
quasimomentum, an equation similar to
 Eq.~(\ref{eq:effEq}) has also been derived
using multiple-scale perturbation theory in the context
of nonlinear optics \cite{sipe88} and atom optics
\cite{lenz94,steel98,kon01}. We have chosen a different approach
since the inclusion of time-dependent quasi momenta is
more obvious using the averaging method.
In the following sections we will apply this equation to examine
the conditions under which gap solitons can be formed and how they
evolve in time.
\section{Derivation of variational equations}
\label{sec:variat}
A major advantage of Eq.~(\ref{eq:effEq}), compared to
the full GPE, is the simple form of the effective Hamiltonians
$H_{\mbox{{\scriptsize  eff}},\|}^{(i)}$. It describes 
interacting particles in a homogeneous external potential
with different masses and velocities. This
allows us to find a simplified analytical description
and thus to gain more insight in the
dynamics of a BEC in an optical lattice. Numerical
simulations of the full GPE indicate that for each
mode the wavepacket remains localized around $q_i$
for a long time if the optical lattice is not too deep.
It is therefore reasonable to assume that the 
wavepackets can approximately be described as Gaussian
wavepackets and to make a variational ansatz for them. 
Following the technique described in 
Refs.~\cite{orgcit2,perez96}, we first observe that 
Eq.~(\ref{eq:effEq}) can formally be derived from
the Lagrangean
\begin{eqnarray} 
  {\cal L} &=&  \sum_i \Big \{ i \frac{\hbar}{2}\left (
   \dot{B}_i^* B_i -B_i^*\dot{B}_i \right )
  \nonumber \\ & &
    + \left ( E(q_i) + \hbar \omega_\perp (n_i+ \frac{1}{2}) 
      - (M a - \dot{q}_i) z \right ) |B_i|^2
  \nonumber \\ & &
    + i \frac{\hbar v_i}{2}\left (
   \partial_z B_i^* B_i -B_i^* \partial_z B_i \right )
   + \frac{\hbar^2}{2  M^{\mbox{{\scriptsize  eff}}}_i}
    |\partial_z B_i|^2 \Big \}
  \nonumber \\ & &
  +\frac{\kappa}{2} \sum_{i,j,k,l} B_i^* B_j^* B_k B_l 
    I^\|_{ij;kl} I^\perp_{ij;kl} .
\end{eqnarray} 
A consistent variational ansatz for Gaussian envelopes is achieved
by setting
\begin{eqnarray}  
  B_i(z,t) &=& \frac{A_i(t)}{\sqrt{\pi^{1/2} w_i(t)}} \exp \Big (
  -\frac{(z-z_i(t))^2}{2 w_i(t)^2} -i \phi_i(t) 
  \nonumber \\ & & \hspace{1cm}
  +i \beta_i(t) z 
  + i \gamma_i(t) z^2 \Big ) .
\label{eq:variatAnsatz}\end{eqnarray} 
This describes a wavepacket of width $w_i$ and amplitude $A_i$ 
(having dimensions of length$^{1/2}$ so that
$B_i$ is dimensionless). It is spatially localized around
$z_i$ and has an instantaneous energy of $\hbar \dot{\phi}_i$.
Its mean velocity and its variance are given by
$\langle v_i \rangle = (\beta_i+ 2 \gamma_i w_i)/
M_i^{\mbox{{\scriptsize  eff}}}$ and
$\Delta v_i = \sqrt{2 \gamma_i^2 w_i^2 + 1/(2 w_i^2)}/
M_i^{\mbox{{\scriptsize  eff}}}$.

Inserting this ansatz for the envelopes in the Lagrangean and
extremizing the corresponding action integral, we derived a set
of 18 equations which describe the evolution of the three
Gaussian wavepackets involved. This task, as well as the algebraic
manipulations following below, are rather tedious and therefore
have been completed using Mathematica \cite{Mathematica}. 
Since the variational equations are somewhat lengthy we exploited
the special features of our system to reduce its complexity.
To do so, we restrict our considerations to the case
when the wavepackets are already at the upper band edge
so that the quasi momenta are time-independent and given by
$q_0 = \hbar k_L$ and $q_1=\hbar k_L -\delta q$ as well as
$q_2=\hbar k_L +\delta q$, where $k_L$ is the wavenumber
of the optical lattice which appears in the optical potential
(\ref{eq:vopt}).
$\delta q$ is identical to $q_2-q_0$. It
can be derived from the resonance condition
that the three energies $E(q_i) + \hbar \omega_\perp (n_i+ \frac{1}{2})$
for $i=0,1,2$ are equal. Setting this energy to zero we can also
omit the corresponding terms in the Lagrangean.
Because the wavepackets are already at the upper band edge we
will also not need the homogeneous force to induce Bloch oscillations,
i.e., we set $a = \dot{q}_i =0$.

The special values of the quasi momenta imply that most of the
interaction integrals $I^\|_{ij;kl}$ are zero or have an identical
value. This can be seen by expanding the
Bloch wavefunctions in terms of momentum eigenstates,
$\varphi_q(z) = \sum_l c_l(q) \exp (i z(q+2 l \hbar k_L)) $.
By Fourier transforming the stationary
Schr\"odinger equation $H_\| \varphi_q = E(q) \varphi_q$,
one finds the following equation for the expansion coefficients
$c_l(q)$,
\begin{equation}
  E(q) c_l(q) = \frac{(q+2l \hbar k_L)^2}{2M} c_l(q) + \frac{V_0}{2}
  \big ( c_{l+1}(q) + c_{l-1}(q) \big ) .
\label{eq:bandEq}\end{equation} 
This equation shows that the expansion coefficients are real and
that, if $c_l(\hbar k_L -\delta q)$ is a solution, then so is
$c_l(\hbar k_L+\delta q) = c_{-l-1}(\hbar k_L-\delta q)$. Thus, we have the 
relation 
\begin{equation} 
  \varphi_{q_2}(z) = \varphi_{q_1}^*(z) \; .
\label{eq:1equal2} \end{equation} 

It is well known, and can be seen from the above expansion,
that Bloch wavefunctions are periodic up to a
phase factor $\exp (i q x)$. Therefore, the three wavefunctions
$\varphi_{q_i}$ are oscillating with a phase factor
$\exp (\pm i \delta q x)$ relative to each other. 
In the limit of an infinite optical lattice, the interaction
integral $I^\|_{ij;kl}$ will therefore vanish if these phase factors
do not exactly cancel each other. For instance, 
$I^\|_{00;01}=0$ because its integrand
is proportional to $\exp ( i \delta q x)$, but
$I^\|_{00;12}\neq 0$. This, in combination with Eq.~(\ref{eq:1equal2}),
ensures that all interaction integrals, except $I^\|_{00;00}$ and
$I^\|_{11;11}=I^\|_{22;22}=I^\|_{12;12}$ as well as
$I^\|_{01;01}=I^\|_{02;02}=I^\|_{00;12}$, do vanish (in addition,
the symmetries $I^\|_{ij;kl}=I^\|_{ji;kl}$ and 
$I^\|_{ij;lk}=I^\|_{ij;kl}$ have to be taken into account).
Thus, there are only three independent interaction parameters
which we will denote by
\begin{eqnarray} 
  \kappa_0 &:=& \frac{\kappa}{\sqrt{\pi}\hbar} I^\|_{00;00} I^\perp_{00;00}\; ,
  \nonumber \\
  \kappa_1 &:=& \frac{\kappa}{\sqrt{\pi}\hbar} I^\|_{11;11} I^\perp_{11;11}\; , 
  \nonumber \\
  \kappa_{01} &:=& \frac{\kappa}{\sqrt{\pi}\hbar} 
            I^\|_{01;01} I^\perp_{01;01} \; . 
\end{eqnarray} 
Even with these simplifications the resulting equations are still
very lengthy, but they admit the analysis of symmetric solutions.
By symmetry, we have $v_0=0$ and $v_2=-v_1$ for the
group velocities of the wavepacket, and 
$ M^{\mbox{{\scriptsize  eff}}}_2= M^{\mbox{{\scriptsize  eff}}}_1$
for the effective masses. Under these conditions one can show that
$z_0 =0$ and $\beta_0 =0$ are solutions of the variational equations.
This result is intuitively clear and just means that the central
wavepacket remains at the upper band edge with mean position and
velocity zero. In addition, symmetry implies that the two transversally
excited wavepackets should evolve in an identical way, but with
opposite mean velocities (because their group velocities differ by
a sign). We therefore can set $A_2=A_1$, $\gamma_2=\gamma_1$,
$\phi_2=\phi_1$, $w_2=w_1$, and $\beta_2=-\beta_1$, $z_2=-z_1$,
which reduces the number of independent variational parameters to ten 
(four for $q_0$ and six for $q_1$). In addition, the
conservation of the number of atoms implies the constraint
$A_0^2 + 2 A_1^2 = L$, so that we are effectively left with
nine independent parameters
\footnote{The amplitudes $A_i$ are normalized to $L$
because the full wavefunction $B_i \varphi_{q_i}$ should be normalized
to one and $\varphi_{q_i}$ carries a factor of $1/\sqrt{L}$ because of
its normalization.}. The resulting variational equations are given by
\begin{equation} 
\dot{A}_0 = \frac{e^{-\frac{z_1^2}{w_1^2}}\,S^{(1)}\,{A_0}\,{{A_1}}^2\,
    {{\kappa }_{01}}}{{w_0}{w_1}} ,
\label{eq:a0dot}\end{equation} 
\begin{equation} 
\dot{z}_1 =
{v_1} + \frac{\hbar \,{{\beta }_1}}{{M^{\mbox{{\scriptsize  eff}}}_1}} + 
  \frac{2\,\hbar \,{z_1}\,{{\gamma }_1}}{{M^{\mbox{{\scriptsize  eff}}}_1}} + 
  \frac{e^{-\frac{z_1^2}{w_1^2}}\,S^{(1)}\,{{A_0}}^2\,{z_1}\,
     {{\kappa }_{01}}}{{w_0}\,{w_1}} ,
\label{eq:z1dot}\end{equation} 
\begin{equation} 
\dot{w}_0 = \frac{2\,\hbar \,{w_0}\,{{\gamma }_0}}{{
   M^{\mbox{{\scriptsize  eff}}}_0}} + 
  \frac{e^{-\frac{z_1^2}{w_1^2}}\,{{A_1}}^2\,
     \left( S^{(3)} - 
       S^{(1)}\,{{w_0}}^2 \right) \,{{\kappa }_{01}}}
     {{{w_0}}^2\,{w_1}} ,
\label{eq:w0dot}\end{equation} 
\begin{eqnarray} 
\dot{w}_1 &=&\frac{2\,\hbar \,{w_1}\,{{\gamma }_1}}{{
  M^{\mbox{{\scriptsize  eff}}}_1}} + 
\label{eq:w1dot} \\ & &
  \frac{e^{-\frac{z_1^2}{w_1^2}}\,{{A_0}}^2\,
     \left( -S^{(3)} + 
       S^{(1)}\,{{w_1}}^2 - 
       2\,S^{(1)}\,{{z_1}}^2 \right) \,{{\kappa }_{01}}
     }{2{w_0}\,{{w_1}}^2} ,
 \nonumber\end{eqnarray}  
\begin{eqnarray} 
\dot{\gamma}_0 &=&
 \frac{\hbar }{2\,{M^{\mbox{{\scriptsize  eff}}}_0}\,{{w_0}}^4} - 
  \frac{2\,\hbar \,{{{\gamma }_0}}^2}{{M^{\mbox{{\scriptsize  eff}}}_0}} + 
  \frac{{{A_0}}^2\,{{\kappa }_0}}
   {2\,{\sqrt{2}}{{w_0}}^3} + 
\label{eq:g0dot} \\ & &
  \frac{e^{-\frac{z_1^2}{w_1^2}}\,{{A_1}}^2\,
     \left( -C^{(3)} + 
       C^{(1)}\,{{w_0}}^2 \right) \,{{\kappa }_{01}}}
     {{{w_0}}^5\,{w_1}} + 
\nonumber \\ & &
  \frac{4\,e^{-\frac{z_1^2}{\bar{w}^2}}\,{{A_1}}^2\,
     \left( {\bar{w}}^2 - 2\,{{z_1}}^2 \right) \,
     {{\kappa }_{01}}}{{\bar{w}}^5 } ,
\nonumber\end{eqnarray} 
\begin{eqnarray} 
\dot{\gamma}_1 &=&
\frac{\hbar }{2\,{M^{\mbox{{\scriptsize  eff}}}_1}\,{{w_1}}^4} - 
  \frac{2\,\hbar \,{{{\gamma }_1}}^2}{{M^{\mbox{{\scriptsize  eff}}}_1}} + 
  \frac{{{A_1}}^2\,{{\kappa }_1}}
   {2\,{\sqrt{2}}{{w_1}}^3} + 
\label{eq:g1dot} \\ & &
  \frac{{e^{-2\frac{z_1^2}{w_1^2}}}\,{{A_1}}^2\,
     \left( {{w_1}}^2 - 4\,{{z_1}}^2 \right) \,{{\kappa }_1}}
     {{\sqrt{2}}{{w_1}}^5} + 
\nonumber \\ & &
  \frac{2{{\kappa }_{01}}\,e^{-\frac{z_1^2}{\bar{w}^2}}\,{{A_0}}^2\,
        \left( {\bar{w}}^2 - 2\,{{z_1}}^2 \right) }{
        {\bar{w}}^5 } + 
\nonumber \\ & &
     \frac{{{\kappa }_{01}}e^{-\frac{z_1^2}{w_1^2}}\,{{A_0}}^2\,
        \left( -C^{(3)} + 
          C^{(1)}({{w_1}}^2 - 
          2{{z_1}}^2) \right) }{2\,
        {w_0}\,{{w_1}}^5} ,
\nonumber\end{eqnarray} 
\begin{eqnarray} 
\dot{\phi}_0 &=&
\frac{\hbar }{2\,{M^{\mbox{{\scriptsize  eff}}}_0}\,{{w_0}}^2} + 
  \frac{5\,{{A_0}}^2\,{{\kappa }_0}}
   {4\,{\sqrt{2}}{w_0}} + 
\label{eq:f0dot} \\ & &
  \frac{e^{-\frac{z_1^2}{w_1^2}}\,{{A_1}}^2\,
     \left( -C^{(3)} + 
       3\,C^{(1)}\,{{w_0}}^2 \right) \,{{\kappa }_{01}}
     }{2\,{{w_0}}^3\,{w_1}} + 
\nonumber \\ & &
  \frac{2\,e^{-\frac{z_1^2}{\bar{w}^2}}\,{{A_1}}^2\,
     \left( 3\,{{w_0}}^4 + 2\,{{w_1}}^4 + 
       {{w_0}}^2\,\left( 5\,{{w_1}}^2 - 2\,{{z_1}}^2 \right) 
       \right) \,{{\kappa }_{01}}}{
     {\bar{w}}^5 } ,
\nonumber \end{eqnarray} 
\begin{eqnarray} 
\dot{\phi}_1 &=&
\frac{\hbar }{2\,{M^{\mbox{{\scriptsize  eff}}}_1}\,{{w_1}}^2} - 
  \frac{\hbar \,{{z_1}}^2}{2\,
  {M^{\mbox{{\scriptsize  eff}}}_1}\,{{w_1}}^4} + 
  {v_1}\,{{\beta }_1} + 
\label{eq:f1dot}\\ & &
\frac{\hbar \,{{{\beta }_1}}^2}
   {2\,{M^{\mbox{{\scriptsize  eff}}}_1}} + 
  \frac{{{\kappa }_2} {{A_1}}^2}{4\,{\sqrt{2 }}{{w_1}}^5}
     \Big( 5\,\big( 1 + 2\,{e^{-2\frac{z_1^2}{w_1^2}}} \big) \,
        {{w_1}}^4 + 
\nonumber \\ & &\hspace{1cm}
     2\,\big( -1 + 2\,{e^{-2\frac{z_1^2}{w_1^2}}}
          \big) \,{{w_1}}^2\,{{z_1}}^2 + 
       16\,{e^{-2\frac{z_1^2}{w_1^2}}}\,{{z_1}}^4 \Big) \,
      + 
\nonumber \\ & &
  \frac{e^{-\frac{z_1^2}{w_1^2}}\,{{A_0}}^2 {{\kappa }_{01}}}{
              4{w_0}\,{{w_1}}^5}
     \big( C^{(3)}(2{{z_1}}^2 - {{w_1}}^2)   + 
\nonumber \\ & & \hspace{1cm}
       C^{(1)} (3{{w_1}}^4 + 
       4{{z_1}}^4) \big) 
     + 
\nonumber \\ & &
  \frac{e^{-\frac{z_1^2}{\bar{w}^2}}\,{{A_0}}^2 {{\kappa }_{01}} 
          }{{\bar{w}}^5}
     \big( 2\,{{w_0}}^4 + 3\,{{w_1}}^4 + 4\,{{z_1}}^4 + 
\nonumber \\ & &\hspace{1cm}
       {{w_0}}^2\,\left( 5\,{{w_1}}^2 + 2\,{{z_1}}^2 \big) 
       \right) ,
\nonumber \end{eqnarray} 
\begin{eqnarray} 
\dot{\beta}_1 &=&
  - \frac{\hbar \,{z_1}}{
  {M^{\mbox{{\scriptsize  eff}}}_1}\,{{w_1}}^4}  - 
  2\,{v_1}\,{{\gamma }_1} - 
  \frac{2\,\hbar \,{{\beta }_1}\,{{\gamma }_1}}{
   {M^{\mbox{{\scriptsize  eff}}}_1}} + 
\label{eq:b1dot} \\ & &
  \frac{{e^{-2\frac{z_1^2}{w_1^2}}}\,{\sqrt{2}}\,
     {{A_1}}^2\,{z_1}\,\left( {{w_1}}^2 + 4\,{{z_1}}^2 \right) \,
     {{\kappa }_1}}{{{w_1}}^5}
   - 
  \frac{{{A_1}}^2\,{z_1}\,{{\kappa }_1}}
   {{\sqrt{2}}{{w_1}}^3} + 
\nonumber \\ & &
  \frac{8\,e^{-\frac{z_1^2}{\bar{w}^2}}\,{{A_1}}^2\,{{z_1}}^3\,
     {{\kappa }_{01}}}{{\bar{w}}^5} + 
\nonumber \\ & &
    \frac{e^{-\frac{z_1^2}{w_1^2}}\,{{A_0}}^2\,{z_1}\,
     \left( C^{(3)} + 
       2\,C^{(1)}\,{{z_1}}^2 \right) \,{{\kappa }_{01}}
     }{{w_0}\,{{w_1}}^5} . 
\nonumber\end{eqnarray} 
In these equations we have introduced the notation
$\bar{w} := \sqrt{w_0^2 + w_1^2}$ and
\begin{eqnarray} 
  S^{(n)} &:=& i  \frac{e^{-2i (\phi_0-\phi_1)}}{
    \left(\frac{1}{w_0^2}+\frac{1}{w_1^2}-2i(\gamma_0-\gamma_1)\right)^{n/2}}
   + c.c. \; ,
\nonumber \\ 
  C^{(n)} &:=&   \frac{e^{-2i (\phi_0-\phi_1)}}{
    \left(\frac{1}{w_0^2}+\frac{1}{w_1^2}-2i(\gamma_0-\gamma_1)\right)^{n/2}}
+ c.c.  \; .
\end{eqnarray} 
The functions $S^{(n)}$ depend on $\phi_i$, $\gamma_i$, and $w_i$
and do vanish for $\phi_1-\phi_0 =0$.

\section{Special solutions of the variational equations}
\label{sec:sol}
{\em Initially empty transverse excited modes:}
A surprising consequence of the variational equations can be seen
immediately: it follows from Eq.~(\ref{eq:a0dot}) that, 
when all atoms are in the central wavepacket ($A_1=0$),
the amplitude $A_0$ and therefore also $A_1$ will not change
in time. Thus, the transversally excited wavepackets would never
be populated. This prediction is a direct consequence of the
assumption $I^\|_{00;01}\; (=I^\|_{00;02}) \;=0$ and in striking
contradiction to the numerical results of Ref.~\cite{kmh02}.
This difference can be resolved when one recalls the conditions
under which our analytical theory is valid.  $I^\|_{00;01}=0$
is exactly fullfilled only in the limit of an infinite optical
lattice. In a finite lattice the fact (discussed above) that
the integrand is oscillating with a phase factor of 
$\exp (\pm i \delta q z)$ only leads to oscillations of $I^\|_{00;01}$,
so that it is zero on average only. Since our wavepackets have a
finite width in quasimomentum space, there will be a finite
excitation probability even when $A_1=0$ initially. In addition,
our theory assumes that the three wavepackets are not overlapping
in quasimomentum space, since only under this condition the
averaging method can yield reasonable results. In practice,
this is not exactly fulfilled and will lead to corrections to
the prediction of the averaged equations. However, the time
scale for transverse excitation out of a central wavepacket
is quite large (typically about 70 ms \cite{kmh02}) so that
the averaged equations should provide a valid description
for shorter times. In fact, the present considerations may provide
another reason for the long time scales for transverse excitations.
In addition, during the preparation of the wavepacket at the upper
band edge through Bloch oscillations, the transversally excited modes
are populated. Therefore, an initial condition with $A_1(0)\neq 0$
is realistic when we describe a system that already  
is prepared at the upper band edge.

On the other hand, when using the initial condition $A_1(0)=0$
we are left with a theory for the central wavepacket only, since
there are never any transversally excited atoms to interact with.
In this case our description reduces to the case considered in
Ref.~\cite{perez96} (but with a negative effective mass) so that
one can transfer most of the results to our case. We therefore
will not discuss it further.

{\em Case of three initial gap solitons:} Another case
of interest is the case when all three wavepackets are initially
forming independent gap solitons. That is, in the absence of mutual
interactions each of the three envelopes corresponds to a
stationary solution of the variational equations with self-interaction.
We can find these solutions by setting $\kappa_{01}=0$ and removing
the terms proportional to $\kappa_1 \exp (-2 z_1^2/w_1^2)$, which
describe the interaction between wavepacket $q_1$ and $q_2$ (see above).
It is easy to see that in this case the soliton solution is given by
$\gamma_i=\beta_i=0$ and $z_1^{\mbox{{\scriptsize  sol}}}= v_1 t$ as well as
\begin{eqnarray} 
  w_i^{\mbox{{\scriptsize  sol}}} &=& 
  - \frac{\sqrt{2}\hbar}{{A^{\mbox{{\scriptsize  sol}}}_i}^2 \kappa_i 
     M^{\mbox{{\scriptsize  eff}}}_i} ,
\label{eq:gapsolsolW} \\ 
  \phi_i^{\mbox{{\scriptsize  sol}}} &=& \phi_i(0) +
  \frac{3 {A^{\mbox{{\scriptsize  sol}}}_i}^2 
    \kappa_i}{4 \sqrt{2}\; w_i^{\mbox{{\scriptsize  sol}}} } t \; .
\label{eq:gapsolsolF} \end{eqnarray} 
The question remains whether this solution is stable against
the presence of the mutual interactions of the three gap solitions.
To answer it, we have made a stability analysis by linearizing the
variational equations in the deviations 
from the soliton solution (\ref{eq:gapsolsolW}), 
(\ref{eq:gapsolsolF}). We set
$w_i = w_i^{\mbox{{\scriptsize  sol}}} + \epsilon \delta w_i $ 
(and similarly for the other dynamical variables) and consider
all equations only to first order in $\epsilon$, whereby the mutual
interaction terms are treated as of first order in $\epsilon$.
This is justified since these terms all include a factor which
exponentially decays in time and thus have limited influence. 
Such a factor arises because the
three wavepackets all have different group velocities and thus separate
after a short time, the exponential being a consequence of the
overlap between the Gaussian wavepackets.
The resulting linearized equations are given by
\begin{equation} 
\dot{\delta A}_0 = e^{-\big (\frac{t\,v_1}{ 
      w_1^{\mbox{{\scriptsize  sol}}}}\big )^2 }
    \frac{2\,\sin (\Delta \phi)
   \,A_0^{\mbox{{\scriptsize  sol}}}\,
    {A_1^{\mbox{{\scriptsize  sol}}}}^2\,{{\kappa }_{01}}}{ 
    w_1^{\mbox{{\scriptsize  sol}}}  } , 
\label{eq:deltaA0}\end{equation} 
\begin{equation} 
  \dot{\delta w}_0 =- {\sqrt{2}}\,{A_0^{\mbox{{\scriptsize  sol}}}}^2\,
    \delta\gamma_0\,{{\kappa }_0}\,
    {w_0^{\mbox{{\scriptsize  sol}}}}^2 ,
\label{eq:deltaW0}\end{equation} 
\begin{equation} 
  \dot{\delta \gamma}_0 =\frac{{A_0^{\mbox{{\scriptsize  sol}}}}^2\,\delta w_0\,{{\kappa }_0}}
   {2\,{\sqrt{2}}\,{w_0^{\mbox{{\scriptsize  sol}}}}^4} + 
  \frac{A_0^{\mbox{{\scriptsize  sol}}}\,\delta A_0\,{{\kappa }_0}}
   {{\sqrt{2}}\,{w_0^{\mbox{{\scriptsize  sol}}}}^3} 
  +  e^{-\big (\frac{t\,v_1}{ 
      w_1^{\mbox{{\scriptsize  sol}}}}\big )^2 } f_{\gamma_0}(t) ,
\label{eq:deltaG0}\end{equation} 
\begin{equation} 
  \dot{\delta \phi}_0 =\frac{-{A_0^{\mbox{{\scriptsize  sol}}}}^2\,\delta w_0\,
       {{\kappa }_0} }{4\,{\sqrt{2}}\,{w_0^{\mbox{{\scriptsize  sol}}}}^2}
   + \frac{5\,A_0^{\mbox{{\scriptsize  sol}}}\,\delta A_0\,{{\kappa }_0}}
   {2\,{\sqrt{2}}\,w_0^{\mbox{{\scriptsize  sol}}}} 
  +  e^{-\big (\frac{t\,v_1}{ 
      w_1^{\mbox{{\scriptsize  sol}}}}\big )^2 } f_{\phi_0}(t) ,
\label{eq:deltaF0}\end{equation} 
\begin{equation} 
  \dot{\delta z}_1 =- \frac{{A_1^{\mbox{{\scriptsize  sol}}}}^2\,
       \left( \delta\beta_1 + 
         2\,t\,{v_1}\,\delta\gamma_1 \right) \,
       {{\kappa }_1}\,w_1^{\mbox{{\scriptsize  sol}}}}{{\sqrt{2}}}
  +  e^{-\big (\frac{t\,v_1}{ 
      w_1^{\mbox{{\scriptsize  sol}}}}\big )^2 } f_{z_1}(t) ,
\label{eq:deltaZ1}\end{equation} 
\begin{equation} 
  \dot{\delta w}_1 =- {\sqrt{2}}\,{A_1^{\mbox{{\scriptsize  sol}}}}^2\,
     \delta\gamma_1\,{{\kappa }_1}\,
     {w_1^{\mbox{{\scriptsize  sol}}}}^2 
  +  e^{-\big (\frac{t\,v_1}{ 
      w_1^{\mbox{{\scriptsize  sol}}}}\big )^2 } f_{w_1}(t) ,
\label{eq:deltaW1}\end{equation} 
\begin{eqnarray} 
  \dot{\delta \beta}_1 &=& - 
  \frac{{{\kappa }_1}\, {v_1} t\,
       \left( \,{A_1^{\mbox{{\scriptsize  sol}}}}^2\,
        \delta w_1\, + 
       2\,A_1^{\mbox{{\scriptsize  sol}}}\,\delta A_1\,
        \,w_1^{\mbox{{\scriptsize  sol}}} 
       \right) }{\sqrt{2}\,{w_1^{\mbox{{\scriptsize  sol}}}}^4}
\nonumber \\ & &
   -2 v_1\delta\gamma_1
  +  e^{-\big (\frac{t\,v_1}{ 
      w_1^{\mbox{{\scriptsize  sol}}}}\big )^2 } f_{\beta_1}(t) ,
\label{eq:deltaB1}\end{eqnarray} 
\begin{equation} 
  \dot{\delta \gamma}_1 =\frac{A_1^{\mbox{{\scriptsize  sol}}}\,{{\kappa }_1}\,
     \left( A_1^{\mbox{{\scriptsize  sol}}}\,\delta w_1 + 
       2\,\delta A_1\,w_1^{\mbox{{\scriptsize  sol}}} \right) }
     {2\,{\sqrt{2}}\,{w_1^{\mbox{{\scriptsize  sol}}}}^4} 
  +  e^{-\big (\frac{t\,v_1}{ 
      w_1^{\mbox{{\scriptsize  sol}}}}\big )^2 } f_{\gamma_1}(t) ,
\label{eq:deltaG1}\end{equation} 
\begin{eqnarray} 
  \dot{\delta \phi}_1 &=& -
  \frac{{A_1^{\mbox{{\scriptsize  sol}}}}^2\,\delta w_1\,
      {{\kappa }_1}\,\left( 2\,t^2\,{{v_1}}^2 + 
        {w_1^{\mbox{{\scriptsize  sol}}}}^2 \right) }
     {4\sqrt{2}\,{w_1^{\mbox{{\scriptsize  sol}}}}^4} 
\nonumber \\ & & +
  \frac{
     A_1^{\mbox{{\scriptsize  sol}}}\,\delta A_1\,
      {{\kappa }_1}\,
      \left( -2\,t^2\,{{v_1}}^2 + 5\,
     {w_1^{\mbox{{\scriptsize  sol}}}}^2 \right) }
     {2\,{\sqrt{2}}\,{w_1^{\mbox{{\scriptsize  sol}}}}^3} 
\nonumber \\ & &    + {v_1}\,\delta\beta_1\,
        +  e^{-\big (\frac{t\,v_1}{ 
      w_1^{\mbox{{\scriptsize  sol}}}}\big )^2 } f_{\phi_1}(t) .
\label{eq:deltaF1}\end{eqnarray} 
The functions $f_\alpha(t)$ depend on the soliton solution parameters
and increase at most polynomially (degree less than 4) in time.
They represent inhomogenities,
similarly to the right-hand side of Eq.~(\ref{eq:deltaA0}).
Because of the exponentially
decaying factor, these terms are only important for times 
$t < w_1^{\mbox{{\scriptsize  sol}}} / v_1$. Therefore, to analyze
the stability of the soliton solution, it is sufficient to solve the
homogeneous linearized equations for a general set of initial conditions,
since for large enough times this correctly describes the general
solution.  

To reduce the length of the linearized equations we have made an
additional approximation.
Our numerical simulations of the full GPE indicate that, 
after the BEC has been transferred to the
upper band edge, the number of atoms in the $q_0$ wavepacket is 
considerably larger than in the other two modes
\footnote{The variational equations presented in this work would
predict that all atoms remain in the transverse ground state since
the excited modes are initially (almost) empty.}. 
Since $A_i^2 = L N_i$,
where $N_i$ is the initial number of atoms in each mode, one can see
that $A_1^{\mbox{{\scriptsize  sol}}} \ll A_0^{\mbox{{\scriptsize  sol}}}$ 
and therefore $w_1^{\mbox{{\scriptsize  sol}}}
\gg w_0^{\mbox{{\scriptsize  sol}}}$. Assuming that this is the case,
we here present the linearized equations only to second order in the ratio
$A_1^{\mbox{{\scriptsize  sol}}}/A_0^{\mbox{{\scriptsize  sol}}}$.

The general solution of the homogeneous linearized equations is not hard to
find. One immediately sees that $\delta A_0$ and therefore, because of
atom number conservation, also $\delta A_1$ are constant in time.
$\delta w_0$ and $\delta \gamma_0$ are then coupled to each other only
so that Eqs.~(\ref{eq:deltaW0}) and (\ref{eq:deltaG0}) are easily
solved. $\delta w_0$ and $\delta \gamma_0$ then generally show a purely
oscillating behaviour. This solution can then be inserted into 
Eq.~(\ref{eq:deltaF0}) 
for the homogeneous phase factor. The latter then grows in time,
in addition to some oscillating factors, proportional to
$ 3 t \kappa_0 A_0^{\mbox{{\scriptsize  sol}}}
 \delta A_0(0)/(\sqrt{2}w_0^{\mbox{{\scriptsize  sol}}}) $. When this
expression is compared to the evolution of the solition phase factor
(\ref{eq:gapsolsolF}) it becomes obvious that this linear increase
in $\delta \phi_0$ just corresponds to a small deviation, proportional to
$\delta A_0(0)/A_0^{\mbox{{\scriptsize  sol}}}$, from the unperturbed
energy of the soliton. We therefore have shown that the central
soliton around quasi momentum $q_0$ is stable against the interaction
with the other two wavepackets since its stability does also not depend
on the evolution of the deviations in these wavepackets.

The situation is quite different for the transversally excited modes.
Repeating the steps leading to the solution for the central wavepacket,
one can see that the solution for $\delta\beta_1$ is given by
\begin{eqnarray}
 \delta\beta_1(t) &=& 
 \delta \beta_1(0) - 2 v_1 t \cos (\Omega_1 t)
     {{\delta \gamma }_1}(0)
 \\ & & 
  -\frac{v_1 t \sin (\Omega_1 t)}{
    {w_1^{\mbox{{\scriptsize  sol}}}}^2 } \left (
   2\frac{\delta A_1(0) }{A_1^{\mbox{{\scriptsize  sol}}}} 
  + \frac{\delta w_1(0)}{w_1^{\mbox{{\scriptsize  sol}}}} \right ) ,
\nonumber
\end{eqnarray} 
with $\Omega_1 := (4/3) d\phi_1^{\mbox{{\scriptsize  sol}}}/dt$.
This growing oscillatory behaviour clearly indicates instability
against any initial deviations $\delta w_1(0), \delta A_1(0),
\delta \gamma_1(0)$, which
unavoidably are introduced by the interaction between the three
wavepackets. 

It is worth to examine the origin of this instability more closely.
Our arguments are based on the fact that the two transversally excited
wavepackets move away from the central wavepacket. This happens because
we have set $\beta_1=\beta_2=0$ for the excited wavepackets, so that
they propagate with the group velocity $\pm v_1$.
Hence, after some time the wavepackets are separated,
so that the mutual interaction disappears and cannot cause instability
anymore. However, setting $\beta_i=0$ in absence of mutual interactions
creates another source of instability: 
even in a strictly one-dimensional situation, a gap (or bright) soliton
with non-vanishing group velocity is only stable 
\footnote{For other values of $\beta_i$ a wavepacket characterized 
by Eqns.~(\ref{eq:gapsolsolW}) and (\ref{eq:gapsolsolF}) is still
a stationary solution of the nonlinear Schr\"odinger equation,
but it is not stable.}
if the phase factor exactly matches the group velocity, 
$\beta_i = - M_i^{\mbox{{\scriptsize  eff}}} v_i/\hbar$ .
Therefore, the instability of the transversally excited wavepackets
is the same as that of an isolated gap soliton with the wrong phase
factor. 

The only possibility to avoid this kind of instability is to
choose the appropriate phase factors
$\beta_2=-\beta_1 = - M_1^{\mbox{{\scriptsize  eff}}} v_1/\hbar$.
As a consequence, the excited wavepackets would remain at their
original position so that the mutual interaction would not
decrease. Since the latter is a resonant coupling between the
three wavepackets a general superposition of three gap solitons
would not correspond to a stationary solution anymore. In the next
section we will demonstrate that for a particular choice of parameters
this problem can be overcome. 
\section{Triple solitons}
A particularly interesting situation appears when one tries
to construct stationary wavepackets which remain spatially localized
around $z_1=0$. As is evident from 
Eq.~(\ref{eq:z1dot}), this is only possible for 
$\beta_1 = - M_1^{\mbox{{\scriptsize  eff}}} v_1/\hbar$. Interestingly,
this condition also guarantees the validity of $\dot{\beta}_1=0$
in Eq.~(\ref{eq:b1dot}), so that this requirement is self-consistent.
The remaining equations will only lead to a stationary solution
if the populations of the three modes are constant, i.e., if
$\dot{A}_0=0$. Apart from the trivial solutions $A_0 =0$ or $A_1=0$
this can be achieved by the condition $S^{(1)}=0$. A natural solution
to this condition is $\phi_1 -\phi_0 =0$ and $\gamma_0=\gamma_1=0$,
whereby the latter assumption also ensures that the widths of the
wavepackets remain constant. A necessary condition for this to hold
are the equations 
\begin{equation} 
  \dot{\phi}_1-\dot{\phi}_0 = \dot{\gamma}_0= \dot{\gamma}_1=0 \; .
\end{equation} 

Using Eqns.~(\ref{eq:g0dot}) - (\ref{eq:f1dot}) 
this leads to algebraic conditions on the widths
and populations of the three modes. 
The simplest way to solve these algebraic conditions
is to, first, fix the ratio between the widths according to
$w_2 = \eta w_1$, where $\eta$ is some positive number.
In addition, we write $\kappa_i = N \bar{\kappa}_i/L$ so that
$\bar{\kappa}_i$ is independent of the total number of atoms $N$
and remains finite when the quantization length $L$ goes to
infinity.  
For these settings we derived solutions of the algebraic conditions
which determine $N$, $w_1$, and the population distribution  
among the modes as a function of $\eta$, $\bar{\kappa}_i$, 
$M_i^{\mbox{{\scriptsize  eff}}}$, and $v_1$.
A particularly nice example is the case when all three wavepackets
have equal width, $w_1=w_0$. The solution then becomes very
compact and is given by
\begin{eqnarray} 
  A_1^2 &=& L
  \frac{3 M^{\mbox{{\scriptsize  eff}}}_1
     \bar{\kappa}_1 - 6 M^{\mbox{{\scriptsize  eff}}}_0
     \bar{\kappa}_{01}         }{
    2 M^{\mbox{{\scriptsize  eff}}}_0
    ( \bar{\kappa}_0 - 3 \bar{\kappa}_{01} )
        + 3 M^{\mbox{{\scriptsize  eff}}}_1 
    ( \bar{\kappa}_1 - 2 \bar{\kappa}_{01} ) } ,
\label{tsa1}\\
  N &=& \frac{2 v_1 \! \left( 
     2 M^{\mbox{{\scriptsize  eff}}}_0
    ( \frac{ \bar{\kappa}_0 }{3} - \! \bar{\kappa}_{01} )  + 
       M^{\mbox{{\scriptsize eff}}}_1 
    ( \bar{\kappa}_1 - \! 2 \bar{\kappa}_{01} ) \right)     }{
    ( 6 \bar{\kappa}_{01}^2 - \bar{\kappa}_0 \bar{\kappa}_1 )
    \sqrt{ 3 M^{\mbox{{\scriptsize  eff}}}_0
      ( M^{\mbox{{\scriptsize eff}}}_1 - M^{\mbox{{\scriptsize eff}}}_0 ) }
     } ,
\label{tsn}\\
  w_1 &=& \frac{ \hbar}{-{M^{\mbox{{\scriptsize  eff}}}_1}\, {v_1}}
  \sqrt{  \frac{{3({M^{\mbox{{\scriptsize  eff}}}_0} 
   - {M^{\mbox{{\scriptsize  eff}}}_1})}}{-2
     {M^{\mbox{{\scriptsize  eff}}}_0}} } ,
\label{tsw1}\end{eqnarray} 
with $w_2 = w_1=w_0$ and $\bar{\kappa}_i := \kappa_i L/N $ 
being independent from the
number of atoms and the quantization length. 

The state characterized by Eqs.~(\ref{tsa1})-(\ref{tsw1}), which we
will refer to as ``triple soliton'', represents
a special coherent superposition of a wavepacket
in the transverse ground state
at the upper band edge of the optical lattice, and two
wavepackets around the transverse resonances. 
The special choice (\ref{tsa1})-(\ref{tsw1}) for the parameters
ensures that the mutual and self-interaction of the wavepackets
exactly cancel the dispersion of each wavepacket due to its
negative effective mass. It also guarantees, within the approximation
that only two resonances are taken into account, that the 
triple soliton does not spread in the transverse direction.
It therefore can be seen as a generalization of the gap soliton
which is unstable against transverse decay. It differs from
the case of a superposition of three gap-soliton wavepackets discussed
above in that the mutual interaction between the wavepackets destroys 
the latter. This is because the stability criterion
(\ref{eq:gapsolsolW}) and (\ref{eq:gapsolsolF}) takes only into account
the self-interaction of each of the three wavepackets.
For the triple soliton the mutual interaction is included as well.

A very interesting feature of the triple soliton is that the width of the
soliton does not depend in any way on the interaction parameters of
the system. It is solely determined by the structure of the lowest
energy band of the optical lattice and in particular is proportional
to the de Broglie wavelength of a particle of mass 
$ - M^{\mbox{{\scriptsize  eff}}}_1$ moving with the velocity $v_1$.
The number $N$ of atoms in the soliton depends on the interaction
parameters, but it vanishes if the group velocity $v_1$ of the
transverse resonances goes to zero, i.e., if the resonances are close
to the band edge. The population of
the three modes depends on the interaction and leads to
consistency requirements: Since $A_0^2$ can only take values between 0
and $L$ we find that the soliton can only exist if the effective masses
fulfill the inequality
\begin{equation} 
  \frac{\bar{\kappa}_0}{3\bar{\kappa}_{01}} \leq 
  \frac{M^{\mbox{{\scriptsize  eff}}}_1}{M^{\mbox{{\scriptsize  eff}}}_0}
  \leq  \frac{2\bar{\kappa}_{01}}{\bar{\kappa}_1} \; .
\label{eq:massCond}\end{equation} 

\begin{figure}
a) \\ \includegraphics[width=6.7cm]{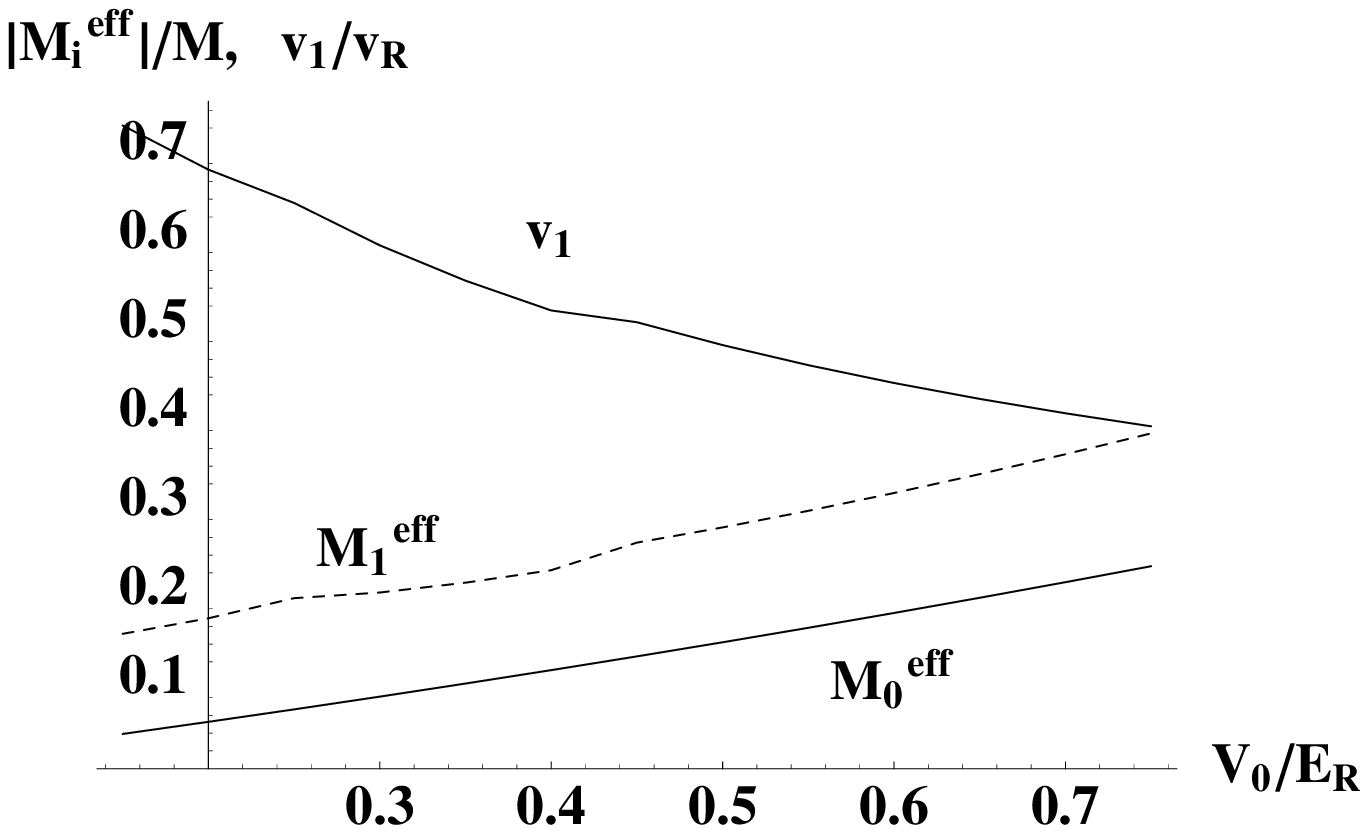}\\
b) \\ \includegraphics[width=6.7cm]{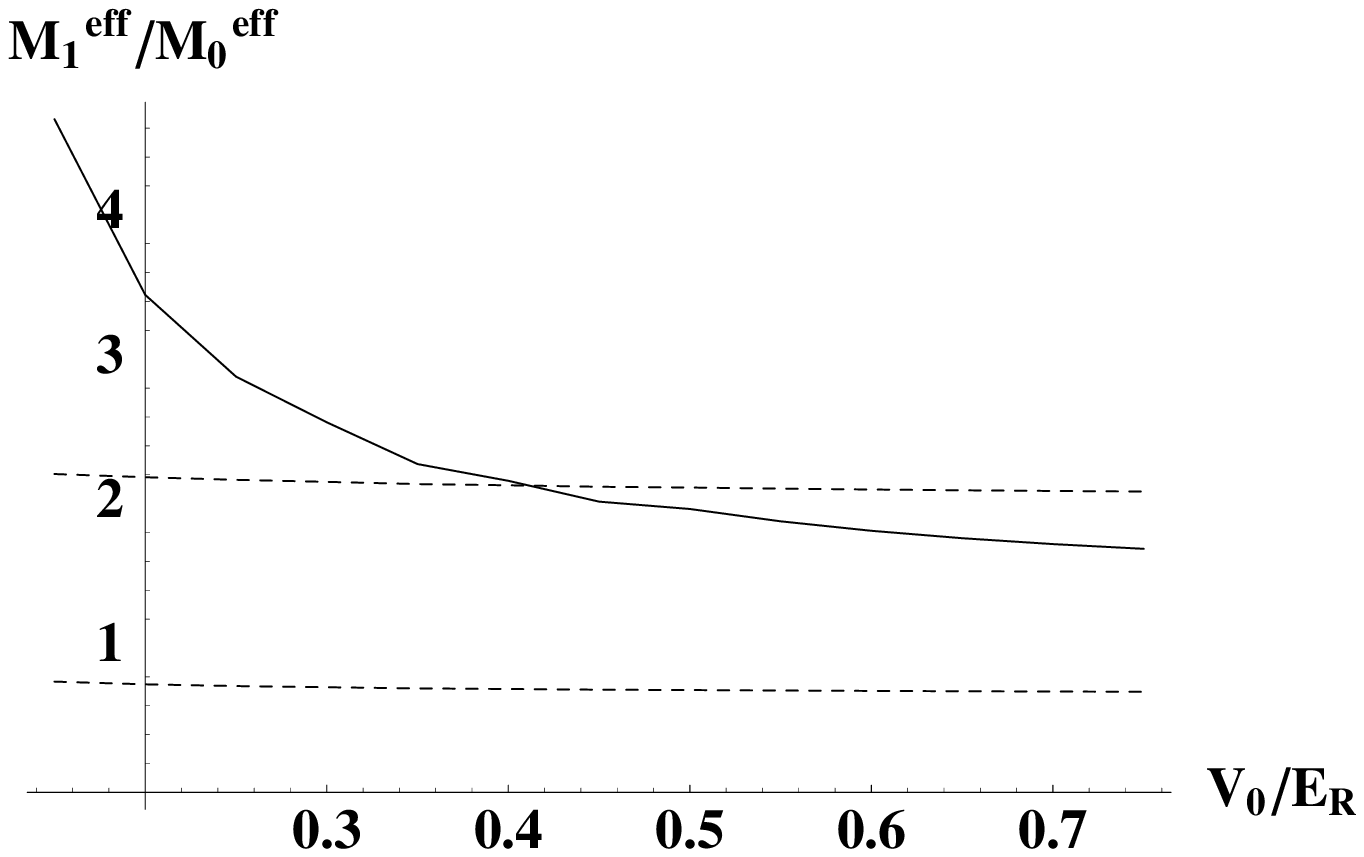}
\caption{\label{fig:effMasses}
a) Group velocity and absolute value of effective masses 
as a function of the optical lattice depth $V_0$.
b) Fullfillment of condition (\ref{eq:massCond}) as a function of $V_0$.
The solid line represents 
$M^{\mbox{{\scriptsize  eff}}}_1/M^{\mbox{{\scriptsize  eff}}}_0$,
the dashed lines are the upper and lower bound in the inequality
(\ref{eq:massCond}). For $V_0 > 0.4 E_R$ the condition is fullfilled.
}
\end{figure}

\begin{figure}
\includegraphics[width=6.7cm]{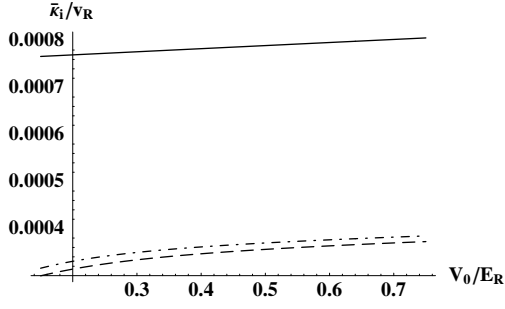}\\
\caption{\label{fig:intParams}
Interaction parameters as a function of lattice depth $V_0$
in units of the recoil velocity $v_R = \sqrt{2E_R /M}$.
Solid line: $\bar{\kappa}_0$, dashed line: $\bar{\kappa}_1$,
dot-dashed line: $\bar{\kappa}_{01}$.
}
\end{figure}

\begin{figure}
a) \\ \includegraphics[width=6.7cm]{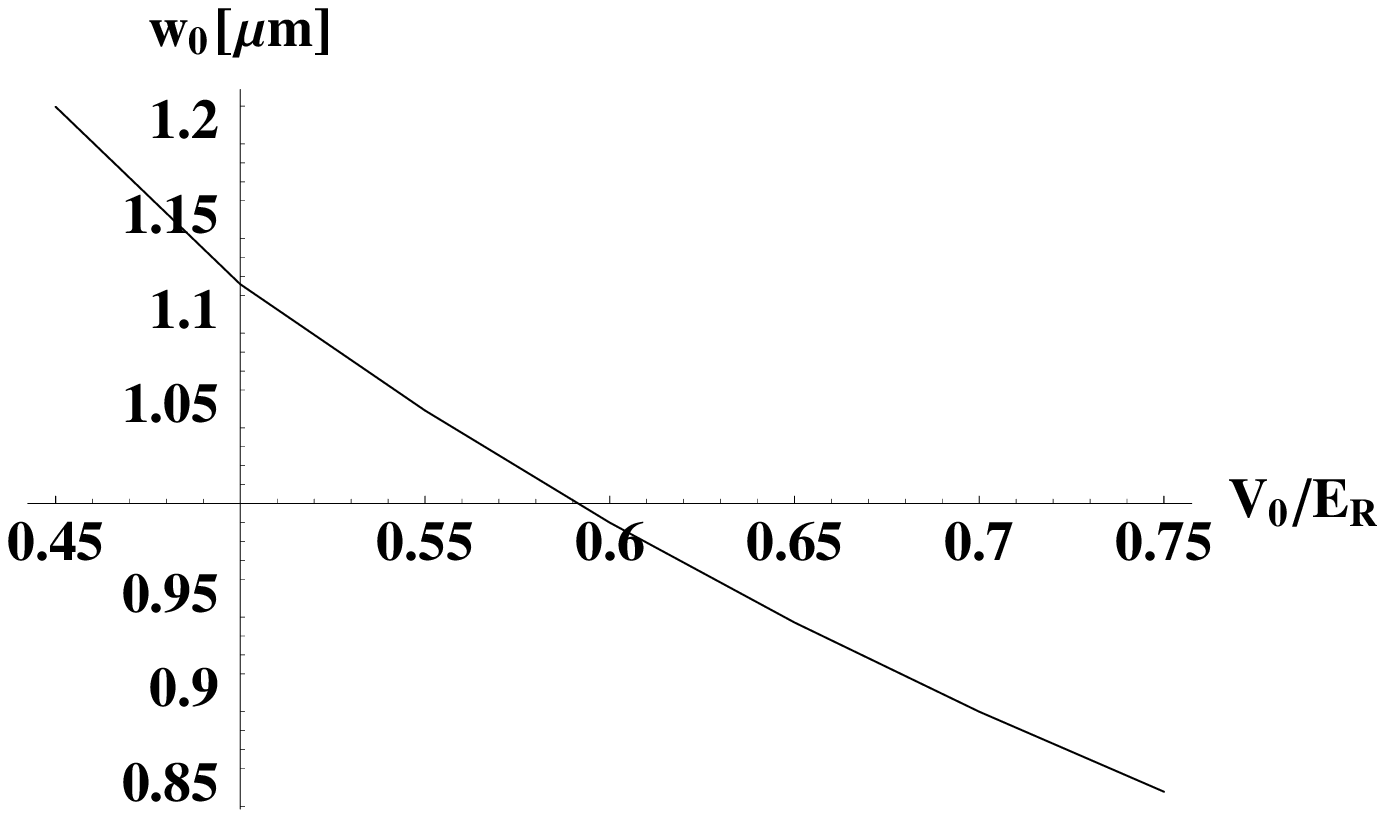}\\
b) \\ \includegraphics[width=6.7cm]{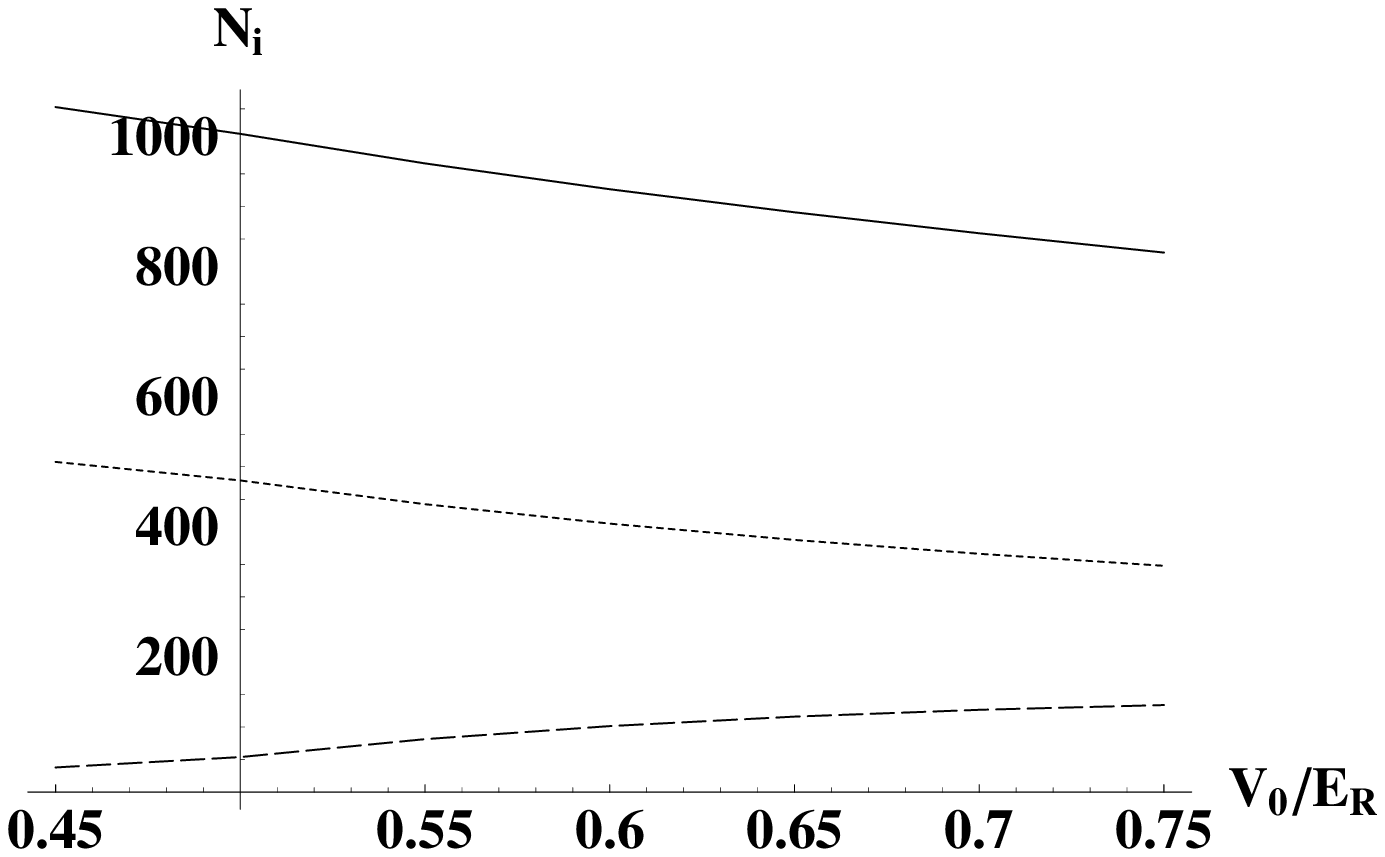}
\caption{\label{fig:solParams}
a) Width of the soliton wavepackets as a function of $V_0$.
b) Total number of atoms $N$ (solid line) in the soliton,
   and number of atoms  $N_i = A_i^2 N/L$ in mode $i=0$ 
(dashed line) and $i=1$ (dotted line), respectively.
}
\end{figure}

To see if this condition can be fullfilled, we have numerically 
calculated the band structure for a BEC in a periodic potential
of the form $ V_0 \cos (2 k_L z )$, where  $k_L=2\pi/\lambda_L$ 
is the laser's wavenumber and $V_0$ the depth of the optical lattice,
which we will give in units of the recoil energy 
$E_R = (1/2) M v_R^2$ with the recoil velocity $v_R = \hbar k_L/M$.
We consider $^{87}$Rb atoms ($M=1.45\times 10^{-25}$ kg, 
$a_{\mbox{{\scriptsize scatt}}}= 4.9 $nm) in an optical lattice
driven by a laser close to the D2 line ($\lambda_L=780$nm)
and a 2-dimensional transverse harmonic trap of strength
$\omega = 534 \mbox{s}^{-1}$. 
The effective mass, the group velocity, and the interaction parameters 
as a function of $V_0$
are shown in Fig.~\ref{fig:effMasses} a) and Fig.~\ref{fig:intParams},
respectively.
As can be seen from  Fig.~\ref{fig:effMasses} b)
condition (\ref{eq:massCond}) can be fulfilled in this parameter regime,
which also lies well within the range of current experiments
\cite{markusPRL03,markusGapSol03}. In Fig.~\ref{fig:solParams}
we have plotted the width as well as the number of atoms and
population distribution for the novel kind of soliton. For the
case $w_0=w_1$ under consideration, the population in the transversally
excited modes is larger than that of the central wavepacket.

\section{Conclusion}

Using an averaging method we have derived effective field equations 
which describe the large-scale 
behaviour of a transversally confined BEC in a
one-dimensional optical lattice. Due to the existence of transversally
excited modes resonant to wavepackets in the transverse
ground state, these equations have the structure of coupled
one-dimensional particles with different effective masses and
dynamical interaction parameters. We have made a Gaussian ansatz
for the envelopes of a wavepacket prepared at the upper band edge
and the two nearest resonances in quasi-momentum space. Variational
equations for this ansatz are derived and several solutions are
discussed, including a novel kind of ``triple'' soliton.

{\bf Acknowledgement}: We wish to thank Markus Oberthaler for many
fruitful discussions. This work was supported
by the Heisenberg-Landau Program, Alberta's informatics Circle of
Research Excellence (iCORE), and the Forschergruppe Quantengase.

One of the authors (V.I.Y.) is grateful to the German Research
Foundation for the Mercator Professorship.


\begin{thebibliography}{99}

\bibitem{orgcit1}
A.S. Parkins and D.F. Walls, Phys. Rep. {\bf 303}, 1 (1998).

\bibitem{orgcit2}
P.W. Courteille, V.S. Bagnato, and V.I. Yukalov, Laser
Phys. {\bf 11}, 659 (2001).

\bibitem{orgcit3}
A.L. Fetter, J. Low Temp. Phys. {\bf 129}, 263 (2002).

\bibitem{orgcit4}
S. Stenholm, Phys. Scr. T {\bf 102}, 89 (2002).

\bibitem{orgcit5}
L. Pitaesvkii and S. Stringari, {\it Bose-Einstein
Condensation} (Oxford University, Oxford, 2003).

\bibitem{castin01} Y. Castin, in {\em Coherent Atomic Matter Waves}, 
  Lecture Notes of Les Houches Summer School, p.1-136, edited by R.
  Kaiser, C. Westbrook, and F. David (EDP Sciences and Springer, Berlin, 
2001).

\bibitem{khayk02} L. Khaykovich, F. Schreck, G. Ferrari, T. Bourdel, 
  J. Cubizolles, L. D. Carr, Y. Castin, and C. Salomon,
  Science {\bf 296}, 1290 (2002).

\bibitem{strecker02} K.E. Strecker, G.B. Partridge, A.G. Truscott, and 
 R.G. Hulet, Nature {\bf 417}, 150 (2002). 

\bibitem{burger99} S. Burger, K. Bongs, S. Dettmer, W. Ertmer, 
  and K. Sengstock, Phys.~Rev.~Lett.~{\bf 83}, 5198 (1999).

\bibitem{denschlag00} J. Denschlag, J.E. Simsarian, D.L. Feder,
  C.W. Clark, L.A. Collins, J. Cubizolles, L. Deng, E.W. Hagley,
  K. Helmerson, W.P. Reinhardt, S.L. Rolston, B.I. Schneider, and
  W.D. Phillips, Science {\bf 287}, 97 (2000).

\bibitem{markusGapSol03} B. Eiermann, T. Anker, M. Albiez, M. Taglieber, 
  P. Treutlein, K.-P. Marzlin, and M.K. Oberthaler,
  Phys.~Rev.~Lett {\bf 92}, 230401 (2004).

\bibitem{markusPRL03}  B. Eiermann, P. Treutlein, Th. Anker, M. Albiez, 
  M. Taglieber, K.-P. Marzlin, and M.K. Oberthaler, 
  Phys.~Rev.~Lett.~{\bf 91}, 060402 (2003).

\bibitem{kivshar04} P.J.Y. Louis, E.A. Ostrovskaya, and Y.S. Kivshar,
   cond-mat/0408291.

\bibitem{kivshar04b} B.J. Dabrowska, E.A. Ostrovskaya and Y.S. Kivshar,
   cond-mat/0408234.

\bibitem{plaja04} L. Plaja and J. San Rom\'an,
    Phys. Rev. A {\bf 69}, 063612 (2004).
\bibitem{konotop04} V.V. Konotop, P.G. Kevrekidis, M. Salerno,
  cond-mat/0404608.
\bibitem{inguscio04} 
  L. Fallani, L. De Sarlo, J. E. Lye, M. Modugno, R. Saers, C. Fort, 
  and M. Inguscio, 
  Phys. Rev. Lett. {\bf 93}, 140406 (2004).

\bibitem{inguscio04b} 
  H. Ott, E. de Mirandes, F. Ferlaino, G. Roati, G. Modugno, and M. Inguscio,
  Phys. Rev. Lett. {\bf 92}, 160601 (2004).

\bibitem{arimondo03} M. Jona-Lasinio, O. Morsch, M. Cristiani, N. Malossi, 
  J. H. M\"uller, E. Courtade, M. Anderlini, and E. Arimondo,
  Phys. Rev. Lett. {\bf 91}, 230406 (2003). 

\bibitem{kmh02}  K.M. Hilligs{\o}e, M.K. Oberthaler, and K.-P. Marzlin, 
  Phys.~Rev.~A.~{\bf 66}, 063605 (2002).

\bibitem{modugno04b} M.~Modugno, C.~Tozzo, and F.~Dalfovo,
  Phys.~Rev.~A {\bf 70}, 043625 (2004).

\bibitem{modugno04} M. Kraemer, C. Menotti, M. Modugno,
   J. Low Temp. Phys. 138, p. 729 (2005).

\bibitem{salasnich02} L.~Salasnich, A.~Parola, and L.~Reatto,
   Phys.~Rev.~A {\bf 65}, 043614 (2002).

\bibitem{salasnich02b} L.~Salasnich, A.~Parola, and L.~Reatto,
   Phys.~Rev.~A {\bf 66}, 043603 (2002).

\bibitem{brazhni04} V.~A.~Brazhnyi, V.~V.~Konotop, and V.~Kuzmiak,
   Phys.~Rev.~A {\bf 70}, 043604 (2004).

\bibitem{trombettoni01} A. Trombettoni and A. Smerzi,  
  Phys. Rev. Lett. {\bf 86}, 2353 (2001).

\bibitem{hakenQFT} H. Haken, {\em Quantum Field Theory of Solids}
(North-Holland, Amsterdam, 1988).

\bibitem{sipe88} J. Sipe and H. Winful, Opt.~Lett.~{\bf 13}, 132
  (1988);
  C.M. de Sterke and J. Sipe, Phys.~Rev.~A {\bf 38}, 5149 (1988);
  C.M. de Sterke and J. Sipe, Phys.~Rev.~A {\bf 39}, 5163 (1989).

\bibitem{lenz94} G. Lenz, P. Meystre, and E.M. Wright,  
        Phys. Rev. A {\bf 50}, 1681 (1994).

\bibitem{steel98} M. Steel and W. Zhang, cond-mat/9810284.

\bibitem{kon01} V.V. Konotop and M. Salerno,
   Phys. Rev. A {\bf 65}, 021602 (2002).

\bibitem{perez96} V.M. P\'erez-Garc\'ia, H. Michinel, J.I. Cirac, 
  M. Lewenstein, and P. Zoller, Phys. Rev. Lett. {\bf 77},
        5320 (1996).

\bibitem{Mathematica} Mathematica 5.0, Wolfram Research, Illinois 2003.

\end{thebibliography}
\end{document}